\documentclass[conference]{IEEEtran}
\IEEEoverridecommandlockouts
\usepackage{cite}
\usepackage{amsmath,amssymb,amsfonts}
\usepackage{algorithmic}
\usepackage{graphicx}
\usepackage{textcomp}
\usepackage{xcolor}
\usepackage{float}
\usepackage{url}
\usepackage{verbatim}
\usepackage{color}

\def\BibTeX{{\rm B\kern-.05em{\sc i\kern-.025em b}\kern-.08em
    T\kern-.1667em\lower.7ex\hbox{E}\kern-.125emX}}
    
\makeatletter
\newcommand{\linebreakand}{%
  \end{@IEEEauthorhalign}
  \hfill\mbox{}\par
  \mbox{}\hfill\begin{@IEEEauthorhalign}
}
\makeatother  
    
\begin{document}

\title{Benefits and Drawbacks of a Graduate Course: An Experience Teaching Systematic Literature Review}

%\begin{comment}
\author{\IEEEauthorblockN{Anderson Y. Iwazaki\\Vinicius dos Santos}
\IEEEauthorblockA{University of São Paulo (ICMC/USP)\\
São Carlos, Brazil \\
iwazaki.anderson@usp.br, 
vinicius.dos.santos@usp.br}
\and
%\IEEEauthorblockN{Vinicius dos Santos}
%\IEEEauthorblockA{University of São Paulo (ICMC/USP)\\
%São Carlos, Brazil \\
%vinicius.dos.santos@usp.br}
%\and
\IEEEauthorblockN{Katia R. Felizardo\\ \'Erica F. de Souza}
\IEEEauthorblockA{Federal University of Technology - Paran\'a (UTFPR)\\
Cornélio Procópio, Brazil \\
katiascannavino@utfpr.edu.br, ericasouza@utfpr.edu.br}
%\and
%\IEEEauthorblockN{\'Erica Ferreira de Souza}
%\IEEEauthorblockA{\textit{dept. name of organization (of Aff.)} \\
%\textit{name of organization (of Aff.)}\\
%City, Country \\
%email address or ORCID}
\linebreakand
\IEEEauthorblockN{Natasha M. C. Valentim}
\IEEEauthorblockA{Federal University of Paraná\\
Curitiba, Brazil \\
natasha@inf.ufpr.br}
\and
\IEEEauthorblockN{Elisa Y. Nakagawa}
\IEEEauthorblockA{University of São Paulo\\
São Carlos, Brazil \\
elisa@icmc.usp.br}
}
%\end{comment}

\maketitle

\begin{abstract}
Graduate subjects (or courses) are commonly offered in graduate courses and can provide specialized knowledge of different topics that are important for the formation of Ph.D. and Master’s students. At the same time, Systematic Literature Review (SLR) has been increasingly adopted in the computing area as a research method to synthesize the state of the art of a given research topic, identify research groups working on that topic, understand the existing limitations and research gaps, and also identify new research directions. However, it is still not well understood the real benefits and drawbacks of offering a subject that addresses SLR for graduate students. Moreover, it is not known the difficulties faced by professors (i.e., educators) to teach this subject. The main goal of this work is to present an experience report of teaching SLR, in particular, the benefits and drawbacks of this subject for computer science graduate students. In addition, this work also presents the essential topics of SLR that we recommend to be taught and a better way to teach them, aiming at supporting graduate courses to offer it. For this, we surveyed computer science graduate students who attended the SLR subject that was taught for almost ten years in our institutions. In particular, we collected the lessons learned, findings, and insights; following, we summarized the benefits and drawbacks for students, the difficulties for professors, and also those essential topics to be taught. As a main result, the SLR subject can be considered a valuable opportunity for graduate students that could use this subject to conduct the required deep literature review of their research topic and have a better comprehension of their research area. Besides and more importantly, this subject can improve important research skills of students, including the ability to recognize research problems, analyze and synthesize data, think critically, and write papers. Therefore, we believe that graduate courses should analyze the possibility of offering the SLR subject.
\end{abstract}

\begin{IEEEkeywords}
Graduate Course, Computer Science, Systematic Literature Review, SLR
\end{IEEEkeywords}

\section{Introduction}
Over the last decade, most research has emphasized the advantages of secondary studies (i.e., Systematic Literature Reviews (SLR) or Systematic Mapping (SM)) compared with traditional reviews~\cite{Kitchenham15, Kuhrmann17, Lavallee13, Riaz10}. %, Felizardo20}.
In particular, an SLR is a type of literature review that systematically collects studies, critically appraises them, and synthesizes findings on a specific research topic~\cite{Kitchenham15}. SLR is also widely known as a valuable means to gather research gaps and topics for further investigation~\cite{Dyba05, Kitchenham11using, Zhang11, Kitchenham15, Niazi15}. 

In parallel, no matter the graduate courses, students need to conduct a deep literature review during their Ph.D. or even Master's projects \cite{Daigneault14, Puljak17}. %, Felizardo20}. 
In this scenario, we observe many computer science graduate students have conducted SLR replacing traditional reviews. For instance, around 50\% of SLR conducted in the Software Engineering (SE) area (of a total of 436 SLR published in SE, within the period from January 2004 to May 2016 \cite{Mendes20} have a Ph.D. student involved). %\cite{Felizardo20}. 
%\m{as a first author}).
Hence, these and other students have benefited from SLR, which enables the identification of the state of the art of their research interest, the existing research gaps, and a better positioning of their project in relation to the others.

At the same time, most computer science graduate courses 
in Brazil and other countries 
have not offered the SLR subject; exceptions are few universities, such as XXX\footnote{To be replaced after blind review.}, YYY\footnote{To be replaced after blind review.}. %University of São Paulo\footnote{\url{https://www5.usp.br/}} (USP) and State University of Campinas\footnote{\url{https://www.unicamp.br}} (UNICAMP). 
Hence, the real benefits and drawbacks of this subject for students are not still entirely understood. It is not also clear how this subject could be structured regarding the taught topics and, more importantly, how to better balance theory and practice. In this scenario, the main issue that motivated the conduction of this work is that many graduate students that could take advantage of this subject may not be benefiting from it.

Regarding the related work, \cite{Oates09} observed the SLR subject could develop research skills in Master's students, including the ability to handle literature and formulate research questions. \cite{Pejcinovic15} pointed out that graduate students should perform SLR to improve their ability to search, select, understand, and critically appraise studies, as well as integrate data and more precisely identify research gaps and improve their academic writing skills. \cite{Borrego15} analyzed how the engineering area has conducted SLR and proposed improvements in the SLR process focusing on engineering graduate courses. In the same direction, \cite{Froyd15} showed the way to adapt the SLR process to be taught in the engineering area. Despite these related works have brought important contributions, as far as we know, there is still a lack of studies investigating the real benefits and drawbacks of SLR subject to graduate students. There are also no recommendations on SLR topics that could be taught and the ways to better teach them.

In this scenario, the main goal of this work is to report an experience teaching SLR, focusing mainly on the benefits and drawbacks of the SLR subject for computer science graduate students. To do that, we performed a long-term study by analyzing the SLR subject offered in the computer science graduate course at XXX\footnote{To be replaced after blind review.}. %USP. 
This one-semester subject that combines theory and practice was offered nine times (once a year), and a total of 153 students attended it. More importantly, SLR was offered as a principal topic, i.e., the subject only focused on SLR and its process. Considering that the design of graduate subjects is sometimes challenging for graduate courses with respect to topics to be included and abilities/knowledge to be improved in students, this work also delineates the essential topics of SLR and experimented means to teach them. Finally, this work intends to show the undeniable value of SLR and also motivate graduate courses to offer it.

The remainder of this work is structured as follows. Section
%\textit{Research Method}
~\ref{Section:ResearchMethod}
presents the research method used in this work. Section
%\textit{Survey Results}
~\ref{sec:survey} 
discusses the survey results. Section
%\textit{Results}
~\ref{Section:results} 
presents the main results of this work, i.e., the benefits and drawbacks of SLR subject to graduate students and the SLR topics and means to teach them. 
%Section ~\ref{Section:discussion} provides a discussion and the threats to the validity. 
Lastly, Section
%\textit{Final Remarks}
~\ref{Section:FinalRemarks} 
provides the final remarks. 

\section{Research Method} \label{Section:ResearchMethod}

\begin{figure}[ht]
    \centering
    \includegraphics[width=0.8
    \linewidth]{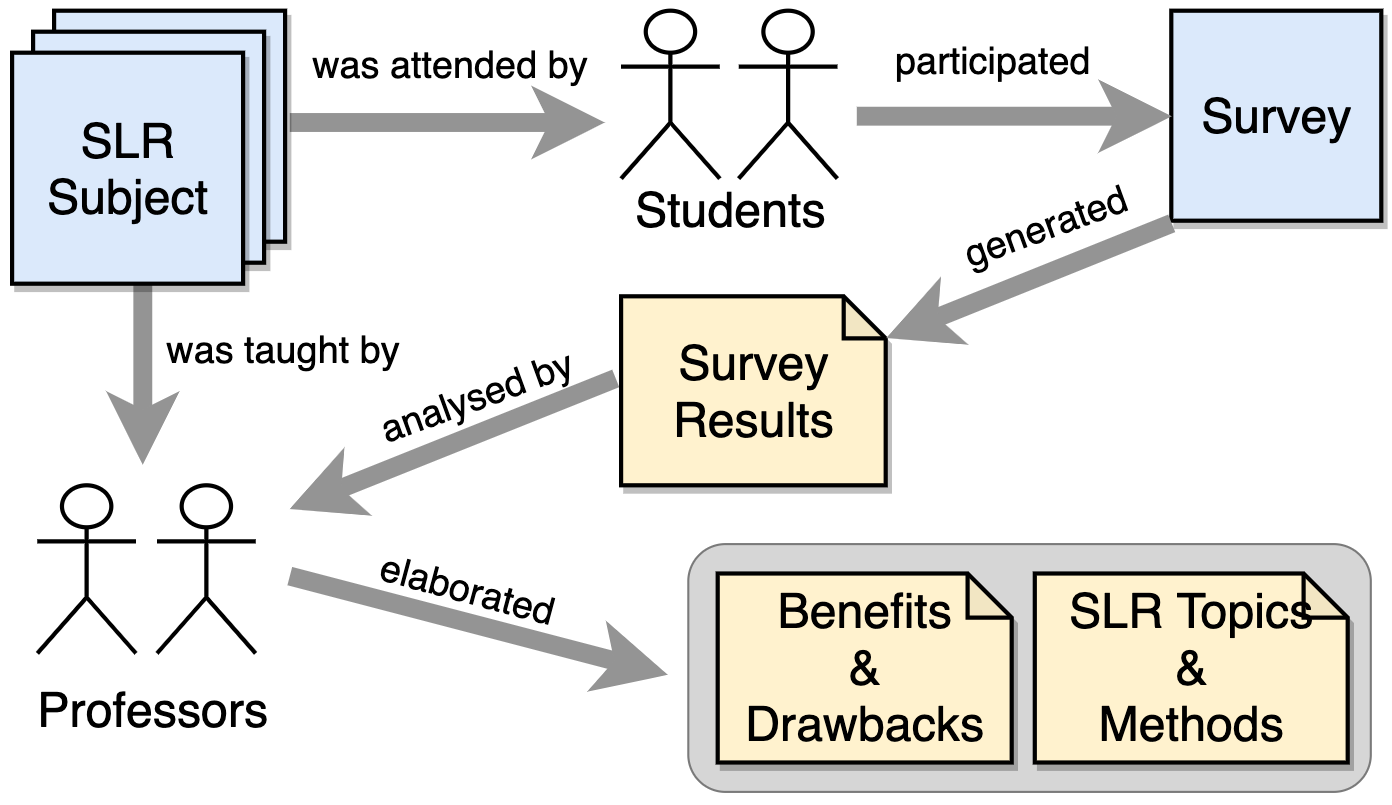}
    \caption{Overall of the research method applied in this work}
    \label{fig:method}
\end{figure}

Figure~\ref{fig:method} summarizes the research method used in this long-term study. It is worth saying three of the authors of this paper are professors (i.e., educators) who have taught the SLR subject in their institutions and have conducted SLR and SM for more than 15 years together with their Ph.D. and Master's students, such as ()\footnote{To be replaced after blind review.}
%\cite{GARCES2017121,GARCES2021111004,8514059, Santos22ESSA, Valle2021}. 
These professors have also worked as researchers on SLR and published their contributions, such as in ()()\footnote{To be replaced after blind review.}. 
%\cite{Nakagawa17, Felizardo17}.

Hence, after these professors taught the SLR subject for almost 10 years, we surveyed students who attended this subject. Following, considering the survey results (presented in Section
%\textit{Survey Results}
\ref{sec:survey}) 
together with our experience, knowledge, and insights collected during the years, we delineated the benefits and drawbacks of this subject for students and also difficulties to professors who teach it (presented in Section %\textit{Benefits, Drawbacks, and Difficulties}
\ref{Section:Benefits}). 
We also delineate the set of topics relevant to be taught as well as better ways to teach them (described in Section %\textit{Topics of SLR Subject}).
\ref{Section:Topics}).
%Next, Sections~\ref{sec:planning} presents the survey planning and conduction.

%\subsection{Survey Planning and Conduction}\label{sec:planning}

In particular, regarding the survey, its planning and conduction were organized in five steps according to \cite{Kitchenham08}:

\begin{itemize}
\item \textbf{Steps 1: Survey definition}: We established four goals (G) to gather the students' opinions: 
\textbf{G1}: Identification of who should attend the SLR subject; 
\textbf{G2}: Identification of the contents that should be taught in the SLR subject; 
\textbf{G3}: Identification of the skills developed in students who attended the SLR subject; and
\textbf{G4}: Identification of difficulties of who attended the SLR subject.
   
\item \textbf{Step 2: Survey instrument development\footnote{
Complete instrument of this survey is available in %\url{https://tinyurl.com/ycmzyogr}}}:
\url{https://drive.google.com/file/d/10lipRCOnUBtuSBNmdjs2-h_aSHTvsT1k/view?usp=sharing}}}: 
The questionnaire had four sections: 
(i) Consent (one question); 
(ii) Student's profile (10 questions); 
(iii) SLR subject (three questions, each one for achieving G1 to G3); and
(iv) Students' impressions (three questions for G4).
Questions associated with G1 to G3 were rated using the Likert scale (Strong agree = 5, Agree = 4, Undecided = 3, Disagree = 2, and Strong Disagree = 1), while those associated with G4 were open questions. 

\item \textbf{Step 3: Evaluation of the survey instrument}: We conducted a pre-test with a smaller sample (a postdoctoral researcher who had already attended the SLR subject) to identify improvements in the questionnaire.
We also analyzed the reliability of the survey considering Cronbach's alpha test \cite{Cronbach51}, which verified that the Likert scale was reliable. The reliability of our survey is 94\% (or Excellent). 

\item \textbf{Step 4: Data collection}: We invited 153 students who attended the SLR subject between 2011 to 2019. They were from various computing areas, including software engineering, artificial intelligence, computing network, human-computer interaction, reconfigurable computing, and education. A total of 32 participants (16 Ph.D. students and 16 Master's students) answered our questionnaire and data was collected through an online form.

\item \textbf{Step 5: Results}: We used two statistical methods (mode and medium) to interpret results associated with G1 to G3. For the open questions (G4), to systematically analyze the information collected from participants, we used Grounded Theory \cite{Seaman08Qualitative, Strauss1998basics}, which encompasses two techniques \cite{Strauss1998basics}: (i) open coding that identifies codes that are separated into discrete parts for analysis; and (ii) axial coding that handles connections between codes and groups them according to their similarities (in our case, the impressions and/or difficulties of students when attending the SLR subject). The next section presents the survey results. 

\end{itemize}

%We also found and mitigated threats to the validity of this survey. Regarding the \textbf{construct validity}, a threat was a given question had influenced the answers to the others. To mitigate it, we concerned with the sentences contained in the questions and ordered the questions to avoid any influence. We also performed a pre-test with a postdoctoral researcher who was also an expert in SLR. In addition, we encouraged respondents to share their impressions through open questions. Concerning the \textbf{conclusion validity}, the representativeness and the number of respondents could be threats. Hence, results cannot be statistically generalized, but we believe our results (i.e., benefits, drawbacks, SLR topics, and methods) are valid because we put together our extensive experience teaching this subject.

\section{Survey Results}\label{sec:survey}

\begin{figure*}[!ht]
    \centering
    \includegraphics[width=0.95
    \linewidth]{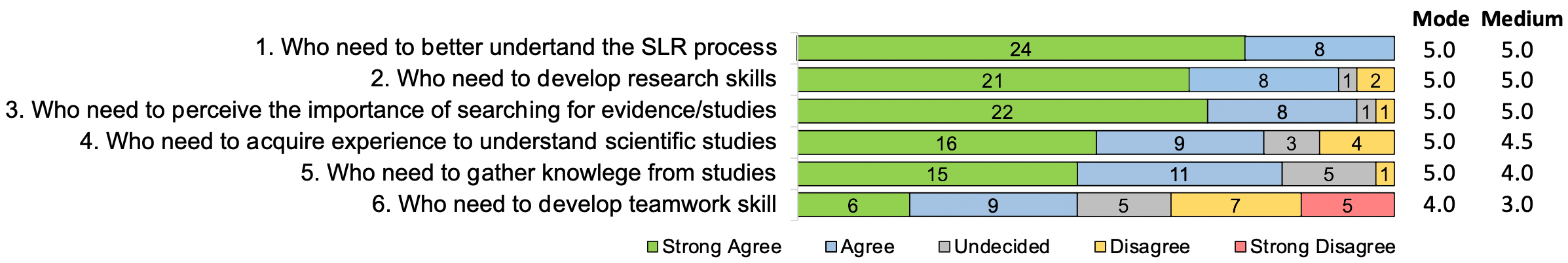}
    \caption{Who should attend the SLR subject}
    \label{fig:g1}
\end{figure*}

Figure \ref{fig:g1} summarizes \textbf{G1} and shows who should attend the SLR subject according to participants\footnote{The raw data is available on %\url{https://tinyurl.com/yyajy9n5}}.
\url{https://drive.google.com/file/d/1_6WYik223__MIOQ-RPSh5X0bVRf5sI-n/view?usp=sharing}}. 
Most of them agreed or strongly agreed with five (of six items), as the mode and medium show. As expected, participants claimed that those who need to understand better the SLR process should attend it. At the same time, students that need to develop research skills (which are actually fundamental for graduate students) should attend it, similarly to those that need to develop the ability to systematically search for evidence/studies in the literature and understand and get knowledge from scientific studies. Otherwise, SLR subject seems not relevant to those that need to develop teamwork skill and networking. We argue this occurs because each student has interest in a specific research topic and, hence, they focus on his/her own SLR; however, students should be closely accompanied by specialists in their research topic (e.g., their supervisors) who should be able to mainly revise confidently the SLR protocol.

\begin{figure*}[ht]
    \centering
    \includegraphics[width=0.8
    \linewidth]{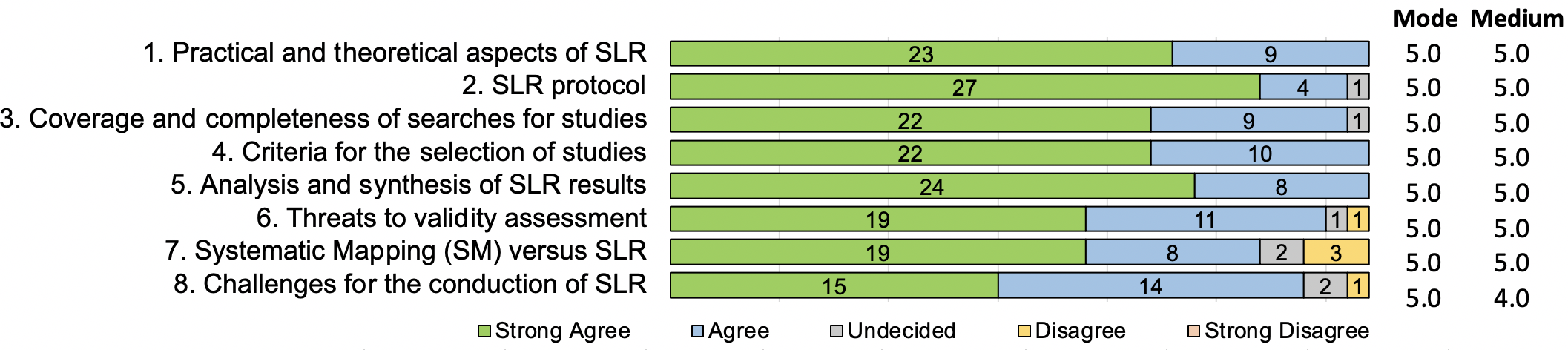}
    \caption{Topics taught in the SLR subject}
    \label{fig:g2}
\end{figure*}

To achieve \textbf{G2}, we questioned participants about the relevance of the topics taught in the subject. Figure \ref{fig:g2} shows that almost all participants agreed or strongly agreed that all topics are relevant, as also pointed out by mode and medium. There is an inclination that the preparation of the SLR protocol and analysis/synthesis of results are more relevant. Meanwhile, the discussion about the difference between SM and SLR as well as challenges to the SLR conduction seem to some extent less relevant compared with the others. We observe all topics considered relevant (i.e., topics 1 to 6) are essential to conducting SLR as a whole, i.e., (i) a good understanding of theoretical and practical aspects; (ii) preparation of a good SLR protocol (which guides the review and documents its execution); (iii) conduction of the SLR following the protocol (i.e., systematic search for studies and application of selection criteria); (iv) synthesis of results; and (v) identification of threats to validity.

\begin{figure*}[!ht]
    \centering
    \includegraphics[width=0.85\linewidth]{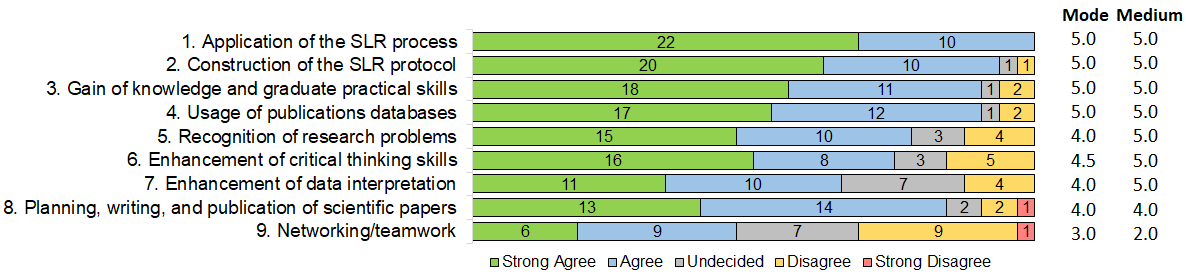}
    \caption{Skills developed in students who attended the SLR subject}
    \label{fig:g3}
\end{figure*}

Concerning \textbf{G3}, Figure \ref{fig:g3} shows most students agreed or strongly agreed that SLR subject made it possible to develop various skills. As expected, two skills or knowledge directly associated with the subject's contents highlighted, i.e., the knowledge about the SLR process and the SLR protocol elaboration. It is also expressive that knowledge and practical skills essential for graduate students are highlighted. Another relevant skill is the ability to accurately use the tools and databases (i.e., database search engines) to find evidence of a given research topic. Other three skills for graduate students can be also developed, i.e., the ability to recognize research problems (of studies being analyzed or even of their graduate projects), the enhancement of thinking critically (on the studies analyzed or even on their studies), and also the ability to interpret data. Another skill considered essential for good researchers is the ability to plan, write, and publish scientific papers. We claim this skill is developed because this subject requires the reading of various studies and also tightly motivates students to publish their SLR in a scientific event or journal. Actually, most respondents (16 of 32) published their SLR in scientific events or specialized journals\footnote{Available on %\url{https://tinyurl.com/yyalotd2}},
\url{https://drive.google.com/file/d/1O5RYzZ5v7FWDbyuFm2lVogbaZbN5ZP5m/view}}, 
while almost all of them (26 of 32) used the SLR in their dissertations/thesis. Otherwise, as previously pointed out, this subject seems not to promote teamwork skills and networking. Moreover, when asked if participants would recommend the subject to their colleagues, they claimed the attendance leveraged their research activities by mainly making it possible to better understand their research area (through the accurate discovery of associated studies), besides the opportunity to publish a paper.

The three open questions gathered the students' impressions and/or difficulties and achieved \textbf{G4}. After analyzing the answers using coding\footnote{Raw data and data analysis are available on %\url{https://tinyurl.com/y2sv5ldw}}
\url{https://drive.google.com/file/d/1ivZ6GxkAHus5FozMdBA8EDtvX0ogWQte/view}}, three issues highlighted: (i) lack of deeper knowledge of their research area; (ii) inherent difficulty of the SLR; and (iii) time constraints. Participants pointed out the \textit{lack of deeper knowledge of their research area} as the main difficulty when attending the subject. Most of them were starting in a new research topic and did not have the confidence to, for instance, define the main goal of the SLR or the right research questions. We observe that almost always students proceed with various changes and adjustments in the SLR protocol, including in the SLR scope, search strategies, and even synthesis of results. This result is in line with a study published in ()\footnote{To be replaced after blind review.}, %\cite{Felizardo17}, 
which showed that the protocol must be iteratively defined through many refinements during the planning phase. These refinements are expected to achieve good SLR, while they can occur independently of the level of experience of the researchers with SLR.

The lack of knowledge of students on their research area directly leads to the \textit{difficulty in all tasks required to the SLR conduction}. Hence, the search for relevant studies was pointed out as challenging due to the need for a suitable search string (composed of keywords, synonymous, and appropriate combination among them). In most cases, students are required to read various studies to identify suitable keywords and their synonyms. The correct use of such string in digital databases is another challenge. In general, these databases are not well-suited to support automated searches; hence, a given search string needs to be accurately adapted to each database, which also continuously updates its search engine (e.g., ACM DL has updated its engine each around two years). Besides, students have sometimes difficulty determining whether a given study is relevant or not only considering the selection criteria. Elaboration of the data extraction form was also mentioned as another difficulty due to the doubt about what to extract from the studies. 

Participants also revealed \textit{time constraint} was another difficulty. Performing an SLR demands considerable time and effort, together with the need to balance effort and time with other activities. Most students start their SLR during the subject, but data extraction and results synthesis are further completed, intending to use the SLR to replace the literature review required for their Master's or Ph.D. projects. 

This survey supported us to the elaboration of the main contributions of this paper,  presented in the next section.

\section{Results}\label{Section:results}

The survey results associated with G1 to G4 together with our experience teaching the SLR subject for years made us possible to find: (i) five benefits of this subject for graduate students; (ii) three drawbacks of this subject for these students; (iii) two main difficulties for professors who teach SLR; and (iv) the main topics that could be taught in the SLR subject and the way to better teach them. 

\begin{figure}[!ht]
    \centering
    \includegraphics[width=1.0\linewidth]{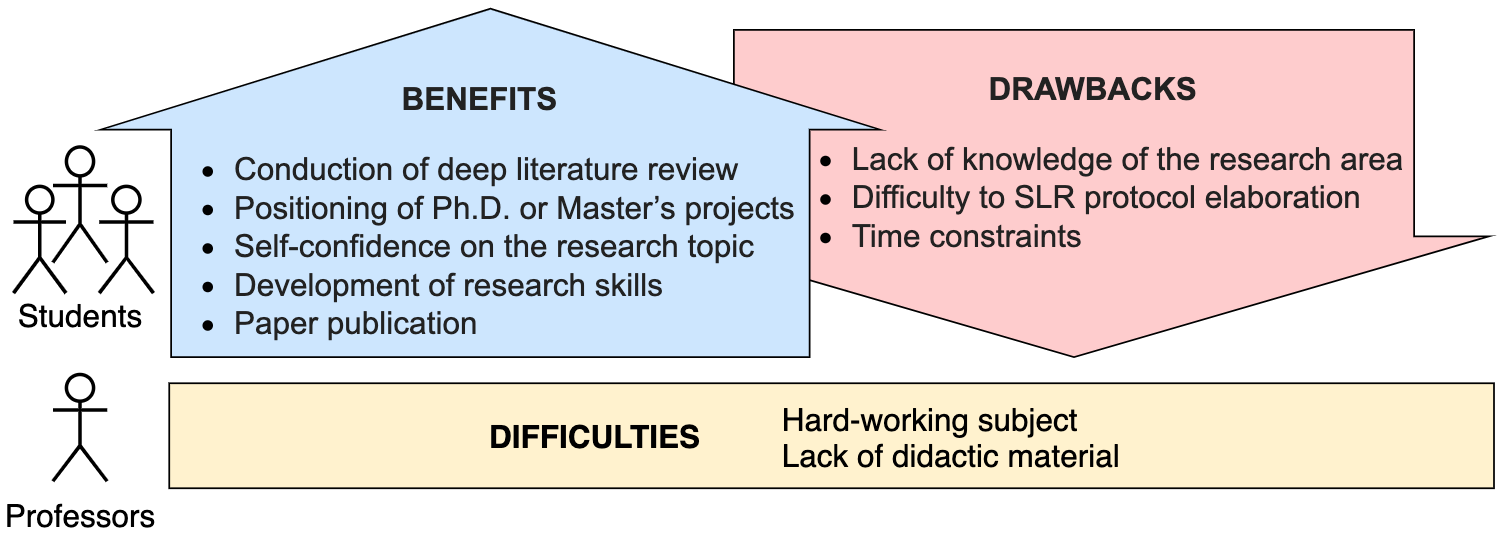}
    \caption{Benefits, Drawbacks, and Difficulties of SLR Subject}
    \label{fig:results}
\end{figure}

%Section~\textit{Benefits, Drawbacks, and Difficulties}
%\ref{Section:Benefits} 
The next subsection
details the benefits, drawbacks, and difficulties, as summarized in Figure~\ref{fig:results}, while the other subsection 
%Subsection~\textit{Topics of SLR Subject}
%\ref{Section:Topics} 
discusses the main topics of SLR.

\subsection{Benefits, Drawbacks, and Difficulties}\label{Section:Benefits}

As with any other graduate subject, the SLR subject can benefits students. We identified a set of five benefits that together could contribute to leverage students to become more prepared researchers in their field.

Considering that graduate students, and undoubtedly all Ph.D. students, should conduct a \textbf{deep literature review} as part of their projects, the SLR subject can provide an opportunity for them to perform that task through a non-biased, transparent, and well-defined process. Grounded on a systematic research method~\cite{Kitchenham15}, a detailed state of the art of the research topic can be gathered, including the identification and combination of commonalities and differences among evidence from various studies, identification of existing research gaps and challenges, and production of reliable and accurate conclusions about a given topic. In this perspective, we recommend students conduct SLR to perform the required deep literature review. However, it is worth highlighting the gains aforementioned can be achieved only if SLR are correctly planned and conducted, following well-experimented guidelines and processes as well as assuring that all tasks are properly performed and auditable. From our experience teaching SLR, we observe there are many decisions that students need to take along with the entire SLR process, and wrong decisions usually lead to many mistakes and, as a consequence, erroneous answers to the research questions. Examples of decisions, just to mention a few, are regarding the: (i) search string (i.e., which keywords, their synonymous, and composition among them should be adopted); (ii) databases (i.e., depending on the research topics, specific databases are required, in particular, when such topics are positioned in a multidisciplinary area, like computing and social science); (iii) use of databases' search engine (i.e., the wrong configuration of the search string for each database's search engine is a recurrent error and, as a consequence, a greater number of not-related studies could be returned or important studies could be missed); 
and (iv) selection criteria (i.e., which the better inclusion and exclusion criteria are, from diverse selection criteria that could be adopted, aiming at correctness and productivity of the selection task). 
At the same time, by usually addressing emerging research topics, almost all SLR conducted during the SLR subject have been unique so that decisions taken in existing SLR even those in closer topics cannot be exactly reused. In this scenario, the SLR subject has supported students to take righter decisions.

Students are also sometimes required to update the literature review to include it in the final version of their theses. In this case, when the review was previously conducted via an SLR, it can be more easily updated. To do that, SLR must have been well-documented, registering not only those elements usually recommended to be documented such as the SLR protocol \cite{Mendes20} but also we particularly recommend documenting those non-trivial elements, such as keywords or databases not used, studies excluded, and all decisions taken during the SLR conduction together with existing alternatives, the selected alternative, and rationales, e.g., all publications databases that were not used and reasons for that.

With their projects ahead still starting, students who have attended the SLR subject do not completely understand yet their projects regarding the exact research problem to be addressed, goals, research method to be followed, and expected contributions. The SLR conduction can support them to identify the state of the art of the associated research, existing gaps, and challenges of their research topic. Hence, another benefit is that the SLR subject makes it possible for students to start to \textbf{better understand and position their Ph.D. or Master's projects} within the research area. At the same time, we have observed the students have recurrently used the SLR results to make decisions about their project, including regarding the scope, goals, and expected contributions. Hence, we believe the SLR subject and the conduction of an SLR are valuable means to drive research, enabling  the opening of novel research topics. It is worth highlighting that students can then conduct evidence-based research, i.e., decisions taken are based on evidence from several studies found in the SLR. More importantly, having a broader and deeper knowledge of the research area, students can use the SLR to somehow assure the originality of their theses.

%\item 
The \textbf{self-confidence of the students on the research topic} is another important benefit. As SLR requires that students systematically and rigorously explore studies in a topic, they become more familiarized with that topic and can gain confidence that they know possibly all related studies. SLR subject could then contribute to reducing the lack of confidence and uncertainty commonly faced by students who are sometimes novices in the area. Moreover, during the SLR conduction, important information about the research topic can be also collected, such as the research tendency (e.g., increase or decrease in the number of studies over the years), main researchers, research groups, and industry involvement working on that topic, leading countries or regions, maturity of the research topic, and research trends (e.g., if the topic and its subtopics are promising). All this knowledge is fundamental for somehow pushing students to become better researchers in their area. 

The fourth benefit is the \textbf{development of research skills} by students. In turn, there exist different sets of research skills desired to be developed in graduate students \cite{Hermanowicz16, Tarasova21, Bravo21, Alhumoud20}. By analyzing these studies, we found the SLR subject could support the development of five important skills:

\begin{itemize}

\item Increase of knowledge on the research topic:
Besides learning a rigorous and effective means to the literature surveys and its associated processes (which are taught in the SLR subject), this subject is an important opportunity for students to increase the knowledge concerning the background and foundations of their research topic by looking into the background section presented in almost all studies. Moreover, students can learn about different scientific methods and assessment methods (presented in the studies), while they can develop practical skills, e.g., dealing better with publications databases;

\item Recognition of research problems: 
SLR subject requires that students look into the existing research problems as well as the research gaps, difficulties, and even contradictions being addressed by studies. Hence, students can develop the ability to recognize them and also the ability to elaborate the research problems for their projects and papers; 

\item Critical thinking: 
The SLR subject faces students to understand different solutions and alternatives, as well as positive and negative points from a variety of studies, pushing these students to engage in reflective and independent thinking and understand the logical connections among different ideas and, as a consequence, enhance the critical thinking;

\item Analysis and synthesis of data: 
SLR subject requires that students summarize data from various studies, define how to organize it, identify relevant data, and synthesize results. All these tasks require the ability to analyze and synthesize data, leading to the development of such important skills; and

\item Writing and publishing: 
SLR subject demands the full reading of various types of scientific literature, such as theoretical and empirical original research, case report, case study, researchers' opinion (or position paper), surveys/reviews, including secondary studies. Students can then accumulate knowledge of this diversity that is surely positive to organize their future publications. At the same time, students can gain experience in writing papers when they report the results of their SLR as a paper.

\end{itemize}

The aforementioned skills are undoubtedly relevant for students, as also discussed in \cite{Froyd15}, mainly those unmeasurable abilities, such as research problem recognition, critical thinking, and paper planning, writing, and publishing. Finally, aligned to the latter (writing and publishing), the fifth benefit is the possibility of students to \textbf{publish a paper} reporting the SLR conduction and results. We observe that, when attending the SLR subject, most students are starting their Ph.D. or Master's projects, which have not still produced results suitable for publication. Hence, by attending this subject, the publication of a paper containing results of their SLR is interesting. As previously found in our survey, around half part of students published the SLR started during the SLR subject in journals and scientific events. At the same time, good SLR are important contributions for any research area \cite{Borrego15}, possibly attracting a number of citations.

Regarding the drawbacks of the SLR subject for students, we identified three drawbacks and also directions that could mitigate them. The first drawback is related to the \textbf{lack of knowledge of students concerning their research area}. Students in their first or second semester of graduate courses have usually attended the SLR subject at XYZ\footnote{To be replaced after blind review.}. %USP. 
At that moment, most of them have not still defined the exact topic of their project and sometimes have little knowledge of the topic that will be possibly researched. However, conducting high-quality SLR requires a deeper knowledge of the research topic \cite{Ribeiro18}. To mitigate this lack, we recommend students have the support of more experienced researchers such as their supervisors to assure the quality of the SLR. If this support is not provided, we have observed students have difficulty proceeding with the steps of the SLR process, e.g., the definition of search strings and elaboration of adequate answers for research questions.

The second drawback refers to the difficulty of the \textbf{SLR protocol elaboration}. The SLR conduction and, in particular, the elaboration of the SLR protocol, are not trivial tasks. Such protocol involves various elements that need to be well-defined (e.g., research questions, search string, search strategies, studies selection process, data extraction form, etc.). To mitigate this difficulty, we recommend building the SLR protocol iteratively, i.e., students perform iterative pre-tests of the protocol to observe its correctness, conciseness, coherence, and completeness and iteratively refine it. Considering our experience from the SLR subject, when such protocol is well-defined by students, they can to some extent more easily perform the next steps of the SLR process.
%%, as also already stated in \cite{Ribeiro18, Felizardo20}. OBS: referências muito locais/regionalizadas!!!!

Another drawback is the \textbf{time constraints} of students. Graduation students have various time- and effort-consuming demands, such as other subjects, a part-time job (in some cases), responsibilities as teaching assistants, and other graduate program requirements that must be completed timely and satisfactorily. In the meantime, the SLR conduction actually consumes considerable time and effort; for instance, it is common for SLR to have more than 2,000 candidate studies to be verified. A big challenge is then to conciliate all academic and non-academic tasks together with the conduction of an SLR. To mitigate this issue, we recommend students focus specifically on the preparation and validation of the SLR protocol during the subject, which usually provides important support for those tasks. Following this, they can start the SLR conduction itself and finish it after the subject. 
We have also observed that the amount of time and effort required for SLR depends directly on the students' skills (e.g., ability of critical thinking and data synthesis) as well as their knowledge of the area. We also have observed that the protocol elaboration can particularly help students to gain such skills and knowledge; hence, the dedication to such protocol becomes even more important.

Concerning the difficulties that professors have faced when teaching SLR, two are relevant to be discussed. The first one is the \textbf{hard-working nature of this subject}. During the subject, each student usually conducts his/her SLR to comply with the literature review necessary for his/her graduate project proposal. In turn, each SLR is sometimes unique in the sense there are no others to be used as a closer reference. In this scenario, professors need to provide meaningful support for each student aiming to make the subject effective. But supporting concomitantly diverse SLR is a very complicated task. It is also hard due to the different rhythm of each SLR with, for instance, different numbers of studies to be managed, data to be extracted, and distinct methods to synthesize results. To minimize this difficulty, a closer involvement of supervisors or other experts in the research topic is necessary, but in many cases, they do not have enough knowledge of particularities of the SLR conduction. At the same time, it is debatable if all students could conduct an SLR of the same research topic, while the professor could have a level of control regarding, e.g., the search string, databases, and selection criteria to be used. We believe such ``toy example'' could be useful only for illustration and exercise, but not for an effective step towards the deep literature review that students need.

Another difficulty is the \textbf{lack of didactic materials}. It is known that there are many scientific materials of SLR (e.g., articles, technical reports, and books) sometimes addressing theoretical issues. However, good didactic materials (in particular, books) are missing besides materials that address practical issues that we believe are core to assure the successful conduction of SLR. Some examples of practical issues are, to mention a few, the configuration of the search string to different databases' search engines, management of sometimes a large number of studies, and correct use of supporting tools. In this scenario, professors have difficulty connecting SLR theory and practice and provide suitable materials to be followed for the students. 

We summarize that as with other subjects, there are difficulties for those who teach the SLR subject and also difficulties for students. However, this subject provides important benefits, in particular, those that can directly contribute to the Ph.D. and Master's theses. Hence, the next section presents some directions of the way that the SLR subject could be taught.

\subsection{Topics of SLR Subject}\label{Section:Topics}

After many adjustments done over the years regarding the topics taught in the SLR subject, we were able to define a set of eight main topics that we believe are comprehensive enough and relevant to be taught, as detailed below: 

\begin{enumerate}

\item Theoretical foundation: This topic presents an overview of various types of literature review, including traditional literature review (a comparison between traditional literature review and SLR is provided in \cite{Zhang11,Zhang13}), rapid review \cite{Khangura12}, multi-vocal literature review \cite{Garousi16,Garousi19}, grey-literature review \cite{Garousi20}, tertiary study \cite{Imtiaz13} and mainly SLR and SM. Comparison among them, highlighting their purposes, differences, and commonalities, as well as advantages and limitations, are discussed;

\item SLR process: 
This topic presents the theoretical aspects of SLR, an overview of the SLR 3-phase process (planning, conduction, and synthesis), and existing guidelines to conduct SLR. More importantly, this topic also encompasses discussions on the iterative nature of the SLR process that provides benefits and difficulties to students who are sometimes conducting their first SLR; 

\item SLR planning: 
This topic addresses the core artifact of SLR, i.e., the SLR protocol. All items of the protocol, including the goal of the SLR, research questions, search string (and the ways to define and calibrate it), search strategy, primary study selection criteria, data extraction methods, and result synthesis methods, are discussed together with the most common approaches to define them;

\item SLR conduction: 
It is presented the means to search studies in the publication databases, the relevant databases for specific areas, and the particularities of each database. This topic also addresses existing strategies to increase the coverage and completeness of the search for studies, such as  experts' opinions and snowballing technique  \cite{badampudi15, jalali12, wohlin14, wohlin16}. Moreover, assuring the coverage and completeness of the searches for studies is another issue presented and widely discussed with students;

\item SLR analysis and synthesis:
This topic presents methods for the data analysis supported by tables or other visual/graphical representations (e.g., graphs, graphics, charts, infographics, word clouds, and geographical maps). Besides that, more importantly, methods for synthesis of SLR results, such as descriptive and quantitative synthesis (obtained through a statistical calculation), are also widely discussed with students;

\item Threats to validity:
As with any other research method, diverse threats to the validity occur in SLR \cite{Ampatzoglou19}. Hence, the main threats to validity, including construct validity, internal validity, and conclusion validity, and also how they could be mitigated, are presented and discussed;

\item SLR update:
In the case of SLR conducted during the SLR subject by graduate students, such SLR are sometimes required to be updated to adhere to the final theses. Hence, this topic discusses the ways and important requirements, mainly regarding the documentation of SLR, so that
SLR are more easily updated \cite{Mendes20}; and

\item Supporting tools: 
Aiming at increasing the productivity and reliability of SLR results, the use of supporting tools is highly recommended. There are tools that can support almost all SLR process such as Parsifal\footnote{\url{https://www.parsif.al}}, while bibliography management tools or reference managers are also useful, e.g., EndNote\footnote{\url{http://endnote.com}},
JabRef\footnote{\url{http://www.jabref.org}}, 
and Mendeley\footnote{\url{https://www.mendeley.com}}. 
\end{enumerate}

It is worth highlighting Topics 1 to 5 should be taught in that order, while the others could be provided along with the semester. For each topic, we also suggest possible ways to teach it, as summarized in Table~\ref{tab:topics} (second column). The first two topics are mainly presented in expository classes, while the remaining are distributed in expository and hands-on classes aiming at achieving the synergy between theory and practice. Moreover, this subject must pay special attention to the learning method adopted to increase the students' engagement and learning. For instance, flipped classroom \cite{Al15, Urfa17} is interesting for this subject since it could leverage the motivation and engagement of students. Another approach is Case-Based Learning (CBL), which uses realistic narratives (cases) for educational purposes \cite{Herreid94CSSN, Yadav07TSCS} and has been already successfully applied in other subjects by promoting the students' learning and their critical thinking \cite{Yadav07TSCS}.

In terms of duration and number of classes, we believe SLR can be adequately addressed in a one-semester subject with around 2-hour class and 6-hour out-of-class per week, whose distribution  is presented in Table~\ref{tab:topics}. Considering a 16-week semester, we believe such distribution of the topics in those classes can work well. In particular, the SLR subject should focus mainly on the SLR planning and conduction consuming approximately 50\% of the classes. As stated before, a deeper discussion on the SLR planning during the classes should be performed, considering the time constraints of students. Regarding the bibliography, there are relevant papers published in main journals and conferences on Empirical Software Engineering that could be adopted in this subject. For each topic, we selected a set of materials that has worked well in our classes, as listed in Table~\ref{tab:topics}. 

As stated before, each SLR developed during the SLR subject has been sometimes unique and has presented its specificity, including its scope, research topic covered, research questions, and publication databases. Although there exists such specificity, some commonality with SLR previously conducted also exists and can be reused. For instance, part of a search string or the same databases used in closer SLR can be used. Hence, good examples of SLR published in important vehicles can serve as references. Good vehicles to find these SLR in the software engineering area are the Journal of Systems and Software\footnote{\url{https://www.journals.elsevier.com/journal-of-systems-and-software}}, Information and Software Technology\footnote{\url{https://www.journals.elsevier.com/information-and-software-technology}}, and conferences like International Conference on Software Engineering (ICSE)\footnote{\url{http://www.icse-conferences.org}}, International Symposium on Empirical Software Engineering and Measurement (ESEM)\footnote{\url{http://www.esem-conferences.org}}, and International Conference on Evaluation and Assessment in Software Engineering (EASE)\footnote{\url{https://conf.researchr.org/home/ease-2022}}.

\begin{table*}[!ht]
\footnotesize
\caption{Essential topics to be taught in the SLR subject}
\label{tab:topics}
\begin{minipage}{1.0\linewidth}
\begin{tabular}{|p{0.13cm}|p{2.1cm}|p{4.6cm}|p{1.2cm}|p{8.0cm} |}\hline
\textbf{\#} & \textbf{Topic} & \textbf{Learning Means} & \textbf{Number of Classes} & \textbf{Main Materials}\\ \hline\hline

1 &
Theoretical foundation & 
Expository class complemented with examples from the literature &
1 (of 16) &
SLR \cite{Kitchenham15}, SM \cite{Kitchenham11using,Petersen15}, traditional literature review \cite{Zhang11,Zhang13}, grey-literature review \cite{Garousi20}, rapid review \cite{Khangura12}, multi-vocal literature review \cite{Garousi16,Garousi19}, and tertiary study \cite{Imtiaz13} \\ \hline

2 &
SLR process & 
Expository class complemented with examples from the literature &
1 (of 16) &
Procedures and guidelines to perform SLR \cite{Kitchenham15} 
\\ \hline

3 &
SLR planning & 
Hands-on classes complemented with expository classes  &
5 (of 16) &
SLR planning \cite{Kitchenham15} 
\\ \hline

4 &
SLR conduction & 
Hands-on classes complemented with expository classes &
3 (of 16) &
SLR conduction \cite{Kitchenham15}, snowballing \cite{badampudi15,jalali12,wohlin14,wohlin16}
\\ \hline

5 &
SLR analysis and synthesis &
Hands-on classes complemented with expository classes &
2 (of 16) &
SLR analysis and synthesis \cite{Kitchenham15} \\ \hline

6 &
Threats to validity & 
Expository class &
1 (of 16) &
Identification, classification, and mitigation of threats to validity in SLR \cite{Ampatzoglou19}
\\ \hline

7 &
SLR update & 
Expository class complemented with examples from the literature &
2 (of 16) &
When and how to update SLR \cite{Mendes20}
\\ \hline

8 &
Supporting tools & 
Hands-on classes &
1 (of 16) &
Practical issues (e.g., configuration of the search string for databases' search engine) and hands-on class using tools such as Parsifal.
\\ \hline

\end{tabular}
\end{minipage}
\end{table*}

%\section{Support for \textsf{\journalclass}}
%We offer on-line support to participating authors. Please contact
%us via e-mail at \dots
%
%We would welcome any feedback, positive or otherwise, on your
%experiences of using \textsf{\journalclass}.

\section{Final Remarks} \label{Section:FinalRemarks}

Graduate subjects can contribute to engaging students in their formation and paves the way to further become them good researchers. This paper presented an experience report on teaching SLR in a subject for computer science graduate students. While diverse benefits and drawbacks for students were observed together with the associated causes, this experience report intends to motivate the discussion about the trade-off to offer this subject. From our point of view, attendance to the SLR subject seems to be a valuable opportunity for students, mainly regarding a means to evolve their Ph.D. or Master's  project. This subject seems to have an important role to develop required research skills in graduate students, including the ability to recognize research problems, analyze and synthesize data, think critically, and write papers. At the same time, professors responsible for this subject have some difficulties teaching it due to the inherent nature of this subject as well as the lack of didactic materials due to SLR is a novel topic to be addressed in graduate courses.

During the conduction of this work, we found and mitigated the threats to the validity. In particular, regarding the survey, a threat to the construct validity was a given question could influence the answers to the others. To mitigate it, we were concerned with elaborating the questions and ordered them to avoid any influence. We also elaborated on all questions considering our long-term experience teaching SLR. Following this, we performed a pre-test with a postdoctoral researcher who was also an expert in SLR. In addition, we encouraged respondents to share their impressions through open questions. Concerning the conclusion validity, the representativeness and the number of respondents could be a threat; hence, we cannot generalize the results, including the benefits, and drawbacks reported in this work, for other graduate courses and students. Different students' backgrounds, the moment when students attend it, and specific issues of the courses can lead to different results compared with ours. Moreover, the difficulties faced by professors who teach this subject cannot be also generalized. Their background and experience with SLR can mitigate or accentuate these or other difficulties. However, we believe that results (i.e., benefits, drawbacks, difficulties, and SLR topics) can be considered relevant because the long-term experience teaching this SLR subject was considered together with the opinion of students that attended the subject along with the years.

We have also observed that SLR has attracted the attention of students from diverse areas, like artificial intelligence, computing network, and education,
%artificial intelligence \cite{Giuntini20}, computing network \cite{Torquato20}, and education \cite{Oliveira18}, OBS: referências muito locais!!!  
besides software engineering that previously introduced SLR in computing. Such areas have also started the publication of SLR and SMS. In this scenario, the SLR subject should deal with specificities of these areas such as the publication databases relevant to be adopted, since they should consider others than those suggested for software engineering. 

For future work, this experience could be replicated in other graduate courses, different regions, and diverse students' profiles. Finally, this work intends to motivate graduate courses not only in software engineering but also in other areas to teach SLR and use it as an added-value means to leverage the students' formation and, as a consequence, the quality of graduate courses.

\begin{comment}
\section*{Acknowledgment}
This work was supported by the Coordination of Superior Level Staff Improvement - CAPES (Grant: PROEX-11357580/D), São Paulo Research Foundation - FAPESP (Grants: 2015/24144-7, 2019/23663-1) and the National Council for Scientific and Technological Development - CNPq (Grants: 312634/2018-8, 313245/2021-5).
\end{comment}

\bibliographystyle{IEEEtran}
\bibliography{bibliografia}

% Generated by IEEEtran.bst, version: 1.14 (2015/08/26)
\begin{thebibliography}{10}
\providecommand{\url}[1]{#1}
\csname url@samestyle\endcsname
\providecommand{\newblock}{\relax}
\providecommand{\bibinfo}[2]{#2}
\providecommand{\BIBentrySTDinterwordspacing}{\spaceskip=0pt\relax}
\providecommand{\BIBentryALTinterwordstretchfactor}{4}
\providecommand{\BIBentryALTinterwordspacing}{\spaceskip=\fontdimen2\font plus
\BIBentryALTinterwordstretchfactor\fontdimen3\font minus
  \fontdimen4\font\relax}
\providecommand{\BIBforeignlanguage}[2]{{%
\expandafter\ifx\csname l@#1\endcsname\relax
\typeout{** WARNING: IEEEtran.bst: No hyphenation pattern has been}%
\typeout{** loaded for the language `#1'. Using the pattern for}%
\typeout{** the default language instead.}%
\else
\language=\csname l@#1\endcsname
\fi
#2}}
\providecommand{\BIBdecl}{\relax}
\BIBdecl

\bibitem{Kitchenham15}
B.~Kitchenham, D.~Budgen, and P.~Brereton, \emph{Evidence-based software
  engineering and systematic reviews}.\hskip 1em plus 0.5em minus 0.4em\relax
  CRC Press, 2015, vol.~4.

\bibitem{Kuhrmann17}
M.~Kuhrmann, ``Teaching empirical software engineering using expert teams,'' in
  \emph{2017 Software Engineering Education in Universities (SEUH)}, 2017, pp.
  20--31.

\bibitem{Lavallee13}
M.~Lavall{\'e}e, P.~Robillard, and R.~Mirsalari, ``Performing systematic
  literature reviews with novices: An iterative approach,'' \emph{IEEE
  Transactions on Education}, vol.~57, no.~3, pp. 175--181, 2013.

\bibitem{Riaz10}
M.~Riaz, M.~Sulayman, N.~Salleh, and E.~Mendes, ``Experiences conducting
  systematic reviews from novices’ perspective,'' in \emph{14th International
  Conference on Evaluation and Assessment in Software Engineering (EASE)},
  2010, pp. 1--10.

\bibitem{Dyba05}
T.~Dyba, B.~A. Kitchenham, and M.~Jorgensen, ``Evidence-based software
  engineering for practitioners,'' \emph{IEEE Software}, vol.~22, no.~1, pp.
  58--65, 2005.

\bibitem{Kitchenham11using}
B.~Kitchenham, D.~Budgen, and P.~Brereton, ``Using mapping studies as the basis
  for further research--a participant-observer case study,'' \emph{Information
  and Software Technology}, vol.~53, no.~6, pp. 638--651, 2011.

\bibitem{Zhang11}
H.~Zhang and M.~Babar, ``An empirical investigation of systematic reviews in
  software engineering,'' in \emph{5th International Symposium on Empirical
  Software Engineering and Measurement (ESEM)}, 2011, pp. 87--96.

\bibitem{Niazi15}
M.~Niazi, ``Do systematic literature reviews outperform informal literature
  reviews in the software engineering domain? {A}n initial case study,''
  \emph{Arabian Journal for Science and Engineering}, vol.~40, no.~3, pp.
  845--855, 2015.

\bibitem{Daigneault14}
P.-M. Daigneault, S.~Jacob, and M.~Ouimet, ``Using systematic review methods
  within a {Ph.D.} dissertation in political science: {C}hallenges and lessons
  learned from practice,'' \emph{International Journal of Social Research
  Methodology}, vol.~17, no.~3, pp. 267--283, 2014.

\bibitem{Puljak17}
L.~Puljak and D.~Sapunar, ``Acceptance of a systematic review as a thesis:
  survey of biomedical doctoral programs in {E}urope,'' \emph{Systematic
  Reviews}, vol.~6, no.~1, pp. 1--8, 2017.

\bibitem{Mendes20}
E.~Mendes, C.~Wohlin, K.~Felizardo, and M.~Kalinowski, ``When to update
  systematic literature reviews in software engineering,'' \emph{Journal of
  Systems and Software}, vol. 167, p. 110607, 2020.

\bibitem{Oates09}
B.~Oates and G.~Capper, ``Using systematic reviews and evidence-based software
  engineering with masters students,'' in \emph{9th International Conference on
  Evaluation and Assessment in Software Engineering (EASE)}, 2009, pp. 1--8.

\bibitem{Pejcinovic15}
B.~Pejcinovic, ``Development and uses of iterative systematic literature
  reviews in electrical engineering education,'' \emph{Electrical and Computer
  Engineering Faculty Publications and Presentations}, vol. 327, no.~1, pp.
  1--10, 2015.

\bibitem{Borrego15}
M.~Borrego, M.~Foster, and J.~Froyd, ``What is the state of the art of
  systematic review in engineering education?'' \emph{Journal of Engineering
  Education}, vol. 104, no.~2, pp. 212--242, 2015.

\bibitem{Froyd15}
J.~Froyd, M.~Foster, J.~Martin, M.~Borrego, H.~Choe, and X.~Chen, ``Special
  session: Introduction to systematic reviews in engineering education
  research,'' in \emph{IEEE Frontiers in Education Conference (FIE)}, 2015, pp.
  1--3.

\bibitem{Kitchenham08}
B.~Kitchenham and S.~Pfleeger, ``Personal opinion surveys,'' in \emph{Guide to
  advanced empirical software engineering}.\hskip 1em plus 0.5em minus
  0.4em\relax Springer, 2008, pp. 63--92.

\bibitem{Cronbach51}
L.~Cronbach, ``Coefficient alpha and the internal structure of tests,''
  \emph{Psychometrika}, vol.~16, no.~3, pp. 297--334, 1951.

\bibitem{Seaman08Qualitative}
C.~Seaman, ``Qualitative methods,'' in \emph{Guide to Advanced Empirical
  Software Engineering}.\hskip 1em plus 0.5em minus 0.4em\relax Springer, 2008,
  pp. 35--62.

\bibitem{Strauss1998basics}
A.~Strauss and J.~Corbin, \emph{Basics of Qualitative Research: Techniques and
  Procedures for Developing Grounded Theory}, 2nd~ed.\hskip 1em plus 0.5em
  minus 0.4em\relax SAGE Publications, 1998.

\bibitem{Hermanowicz16}
J.~Hermanowicz, ``Faculty perceptions of their graduate education,''
  \emph{Higher Education}, vol.~72, no.~3, pp. 291--305, 2016.

\bibitem{Tarasova21}
E.~Tarasova, O.~Khatsrinova, G.~Fakhretdinova, and A.~Kaybiyaynen,
  ``Project-based learning activities for engineering college students,''
  \emph{Advances in Intelligent Systems and Computing}, vol. 1329, pp.
  253--260, 2021.

\bibitem{Bravo21}
P.~Bravo and M.~de~la Rosa, ``Research competences in university training [las
  competencias investigadoras en la formación universitaria],''
  \emph{Universidad y Sociedad}, vol.~13, no.~1, pp. 17--25, 2021.

\bibitem{Alhumoud20}
S.~Alhumoud, A.~Alowisheq, and N.~Altwairesh, ``Teaching research methods for
  computer science students using virtual learning: A case study,'' in
  \emph{12$^{th}$ International Conference on Computer Supported Education
  (CSEDU'20)}, vol.~1, 2020, pp. 518--522.

\bibitem{Ribeiro18}
T.~V. Ribeiro, J.~Massollar, and G.~H. Travassos, ``Challenges and pitfalls on
  surveying evidence in the software engineering technical literature: an
  exploratory study with novices,'' \emph{Empirical Software Engineering},
  vol.~23, no.~3, pp. 1594--1663, 2018.

\bibitem{Zhang13}
H.~Zhang and M.~Babar, ``Systematic reviews in software engineering: An
  empirical investigation,'' \emph{Information and Software Technology},
  vol.~55, no.~7, pp. 1341--1354, 2013.

\bibitem{Khangura12}
S.~Khangura, K.~Konnyu, R.~Cushman, J.~Grimshaw, and D.~Moher, ``Evidence
  summaries: the evolution of a rapid review approach,'' \emph{Systematic
  Reviews}, vol.~1, no.~10, pp. 1--9, 2012.

\bibitem{Garousi16}
V.~Garousi, M.~Felderer, and M.~M{\"a}ntyl{\"a}, ``The need for multivocal
  literature reviews in software engineering: complementing systematic
  literature reviews with grey literature,'' in \emph{20$^{th}$ International
  Conference on Evaluation and Assessment in Software Engineering (EASE)},
  2016, pp. 1--6.

\bibitem{Garousi19}
V.~Garousi, M.~Felderer, and M.~M\"{a}ntyl\"{a}, ``Guidelines for including
  grey literature and conducting multivocal literature reviews in software
  engineering,'' \emph{Information and Software Technology}, vol. 106, no.~1,
  pp. 101--121, 2019.

\bibitem{Garousi20}
------, ``Benefiting from the grey literature in software engineering
  research,'' in \emph{Contemporary Empirical Methods in Software Engineering},
  M.~Felderer and G.~Travassos, Eds.\hskip 1em plus 0.5em minus 0.4em\relax
  Springer, 2020, pp. 389--418.

\bibitem{Imtiaz13}
S.~Imtiaz, M.~Bano, N.~Ikram, and M.~Niazi, ``A tertiary study: Experiences of
  conducting systematic literature reviews in software engineering,'' in
  \emph{EASE}, 2013, p. 177–182.

\bibitem{badampudi15}
D.~Badampudi, C.~Wohlin, and K.~Petersen, ``Experiences from using snowballing
  and database searches in systematic literature studies,'' in \emph{19th
  International Conference on Evaluation and Assessment in Software Engineering
  (EASE)}, 2015, pp. 1--17.

\bibitem{jalali12}
S.~Jalali and C.~Wohlin, ``Systematic literature studies: database searches vs.
  backward snowballing,'' in \emph{6th International Symposium on Empirical
  Software Engineering and Measurement (ESEM)}, 2012, pp. 29--38.

\bibitem{wohlin14}
C.~Wohlin, ``Guidelines for snowballing in systematic literature studies and a
  replication in software engineering,'' in \emph{18th International Conference
  on Evaluation and Assessment in Software Engineering (EASE)}, 2014, p.~38.

\bibitem{wohlin16}
------, ``Second-generation systematic literature studies using snowballing,''
  in \emph{20$^{th}$ International Conference on Evaluation and Assessment in
  Software Engineering (EASE'16)}, 2016, pp. 1--15.

\bibitem{Ampatzoglou19}
A.~Ampatzoglou, S.~Bibi, P.~Avgeriou, M.~Verbeek, and A.~Chatzigeorgiou,
  ``Identifying, categorizing and mitigating threats to validity in software
  engineering secondary studies,'' \emph{Information and Software Technology},
  vol. 106, pp. 201--230, 2019.

\bibitem{Al15}
A.~Al-Zahrani, ``From passive to active: The impact of the flipped classroom
  through social learning platforms on higher education students' creative
  thinking,'' \emph{British Journal of Educational Technology}, vol.~46, no.~6,
  pp. 1133--1148, 2015.

\bibitem{Urfa17}
M.~Urfa and G.~Durak, ``Implementation of the flipped classroom model in the
  scientific ethics course.'' \emph{Journal of Education and e-Learning
  Research}, vol.~4, no.~3, pp. 108--117, 2017.

\bibitem{Herreid94CSSN}
C.~Herreid, ``Case studies in science - a novel method of science education,''
  \emph{Journal of College Science Teaching}, vol.~23, no.~4, pp. 221--229,
  1994.

\bibitem{Yadav07TSCS}
A.~Yadav, M.~Lundeberg, M.~DeSchryver, K.~Dirkin, N.~Schiller, K.~Maier, and
  C.~Herreid, ``Teaching science with case studies: A national survey of
  faculty perceptions of the benefits and challenges of using cases,''
  \emph{Journal of College Science Teaching}, vol.~37, no.~1, pp. 34--38, 2007.

\bibitem{Petersen15}
K.~Petersen, S.~Vakkalanka, and L.~Kuzniarz, ``Guidelines for conducting
  systematic mapping studies in software engineering: An update,''
  \emph{Information and Software Technology}, vol.~64, pp. 1--18, 2015.

\end{thebibliography}

\end{document}